\documentclass[aps,prb,reprint,amsmath,amssymb,superscriptaddress,twocolumn,nofootinbib]{revtex4-2}
\usepackage{graphicx}
\usepackage{xcolor}
\usepackage{subfigure}
\usepackage{hyperref,hypcap}
\usepackage{braket}
\usepackage{amsmath}
\usepackage{commath}
\usepackage{mathtools}
\usepackage{chemformula}

\begin{document}
	
	\title{Intrinsic staggered spin-orbit torque for the electrical control of antiferromagnets - application to \ch{CrI3}}
	
	\author{Fei Xue} 
	\affiliation{Physical Measurement Laboratory, National Institute of Standards and Technology, Gaithersburg, MD 20899, USA}
	\affiliation{Institute for Research in Electronics and Applied Physics \& Maryland Nanocenter,	University of Maryland, College Park, MD 20742, USA}
	\affiliation{Department of Physics,	University of Alabama at Birmingham, Birmingham, AL 35294, USA}	
	\author{Paul M. Haney}	
	\affiliation{Physical Measurement Laboratory, National Institute of Standards and Technology, Gaithersburg, MD 20899, USA}
	
	\date{\today}
	
	\begin{abstract}
		Spin-orbit torque enables the electrical control of the orientation of ferromagnets' or antiferromagnets' order parameter.  In this work we consider antiferromagnets in which the magnetic sublattices are connected by inversion+time reversal symmetry, and in which the exchange and anisotropy energies are similar in magnitude.  We identify the staggered dampinglike spin-orbit torque as the key mechanism for electrical excitation of the N\'eel vector for this case.  To illustrate this scenario, we examine the 2-d Van der Waals antiferromagnetic bilayer \ch{CrI3}, in the $n$-doped regime.  Using a combination of first-principles calculations of the spin-orbit torque and an analysis of the ensuing spin dynamics, we show that the deterministic electrical switching of the N\'eel vector is the result of dampinglike spin-orbit torque which is staggered on the magnetic sublattices.
	\end{abstract}
	
	\maketitle
	
	\section{Introduction}
	Spin-orbit torque is a mechanism for electrically switching thin-film magnets, and has the potential to enable scalable magnetic random access memory and devices for next-generation computing~\cite{manchon2019current}. The effect occurs in magnetically ordered systems that lack inversion symmetry - such as heavy metal-ferromagnet bilayers~\cite{liu2011spin,miron2011perpendicular} - when a DC current or an electric field is applied. Spin-orbit torque can be decomposed into a component that is even under time-reversal, which is also known as the ``dampinglike'' torque, and a component that is odd under time-reversal, known as the ``fieldlike'' torque~\cite{footnote0}.  Knowledge of the dominant component of spin-orbit torque can help to identify the microscopic source of the torque and assist in optimizing the effect~\cite{go2020theory}.  
	
	In addition to switching ferromagnets, spin-orbit torque has been shown to switch antiferromagnets~\cite{Gomonay2010,Gomonay2012,Zelezny2014,wadley2016electrical,Jungwirth2016,Zelezny2017,Mn2Au_Experiment2018,AFM_RMP_2018,Godinho2018}.  Antiferromagnets are of particular interest due to their insensitivity to stray magnetic fields and the fast time scales of their excitations~\cite{Jungwirth2016,AFM_RMP_2018,manchon2019current}.  It was shown~\cite{Zelezny2014} that spin-orbit torque is present in bulk antiferromagnets in which inversion symmetry is {\it locally} broken on individual magnetic sublattices, while the crystal lattice retains global inversion symmetry.  More precisely, in antiferromagnets that are invariant under the combined operations of inversion and time-reversal, the spin-orbit torques acting on the magnetic sublattices can re-orient the antiferromagnetic N\'eel vector ${\bf L}$~\cite{Zelezny2014,wadley2016electrical}.  For previously studied materials with this symmetry, such as CuMnAs and \ch{Mn2Au}, the magnetic exchange energy is much larger than other energy scales, and the mechanism for switching is a uniform fieldlike torque present on both magnetic sublattices \cite{Zelezny2014,wadley2016electrical}.
	
	In this work we focus on a different mechanism for the electrical excitation of the N\'eel order: a staggered dampinglike torque. This torque competes directly with the exchange torque and has therefore been neglected in previous studies, where the exchange torque dominates.  However, there are several materials in which the exchange and anisotropy energies are comparable.  Examples of such materials include \ch{MnPSe3}~\cite{Jeevanandam1999,ni2021imaging}, and chromium trihalides~\cite{ChromiumTrihalides}. We focus on bilayer \ch{CrI3}~\cite{Huang2017} and show that the staggered dampinglike torque can play the dominant role in the spin-orbit torque switching. 
	
	This paper is organized as follows: In Sec.~\ref{Sec:general}, we present a stability analysis of antiferromagnetic spin dynamics for different configurations of spin-orbit torque. We show that staggered dampinglike torque efficiently excites antiferromagnet dynamics when the exchange and anisotropy energies are comparable.  To illustrate this behavior we consider a specific example of bilayer \ch{CrI3}.   In Sec.~\ref{Sec:CrI3 dynamics}, we examine the symmetry properties of \ch{CrI3} and show that the magnetic dynamics occur within a subspace of magnetic configurations. In Sec.~\ref{Sec:CrI3 SOT results}, we present first-principles calculations of spin-orbit torques in \ch{CrI3}.  In Sec.~\ref{Sec:CrI3 LLG results}, we plug these microscopically-computed spin-orbit torques the into Landau-Lifshitz-Gilbert equation to numerically demonstrate electrical switching {\it n}-doped \ch{CrI3} via staggered dampinglike torque. In Sec.~\ref{Sec:Discussion}, we discuss the experimental implications of our main findings.

	\section{Spin-orbit torque in collinear antiferromagnets}
	\label{Sec:general}
	We first consider the spin dynamics of antiferromagnetically coupled spins with various forms of spin-orbit torques.  The time evolution of the spin orientations $\hat{\mathbf{m}}^{\rm A,B}$ are described by the coupled set of Landau-Lifshitz-Gilbert (LLG) equations \cite{Stiles2006,Gomonay2010,Manchon2017}:
	\begin{eqnarray}
		\label{eq:LLG}
		\frac{d{\hat{\mathbf{m}}}^{\rm A,B}}{dt}&=&\hat{\mathbf{m}}^{\rm A,B} \times \left(\frac{\gamma}{m}\frac{\delta E}{\delta \hat{\mathbf{m}}^{\rm A,B}}  +\alpha~\frac{d{\hat{\mathbf{m}}}^{\rm A,B}}{dt}\right)+\boldsymbol{\mathcal{T}}^{\rm A,B},\nonumber\\
	\end{eqnarray}
	where $m$ is the magnitude of the magnetic moment (assumed equal on both sublattices), $\gamma$ is the absolute value of the gyromagnetic ratio, and $\alpha$ is the Gilbert damping parameter.  The energy $E$ is comprised of an easy-axis anisotropy (along $\hat{\bf z}$) and Heisenberg exchange coupling: $E(\hat{\mathbf{m}}^{\rm A},\hat{\mathbf{m}}^{\rm B})=-\frac{1}{2}m H_A[\left(\hat{\mathbf{m}}^{\rm A}\cdot\hat{\bf z}\right)^2+\left(\hat{\mathbf{m}}^{\rm B}\cdot\hat{\bf z}\right)^2 ] + m H_E\left(\hat{\mathbf{m}}^{\rm A}\cdot\hat{\mathbf{m}}^{\rm B}\right)$, where $H_A$ and $H_E$ are the effective magnetic fields from anisotropy and exchange, respectively. $\boldsymbol{\mathcal{T}}^{\rm A,B}$ is the spin-orbit torque on the A,B sublattice. As mentioned earlier, the torque is classified as either even or odd under time-reversal, and we additionally distinguish between torques which are equal or opposite in sign on the two sublattices (denoted ``uni''form or ``stagg''ered). These combinations result in four independent contributions to the spin-orbit torque summarized in Table~\ref{Table:SOT}.
	
	\begin{table}[htbp]
		\label{Table:SOT}
		\centering
		\begin{tabular}{ |p{17.5mm}|c|c| } 
			\hline
			& $\boldsymbol{\mathcal{T}^{\rm A}}$ & $\boldsymbol{\mathcal{T}^{\rm B}}$ \\ 
			\hline
			staggered\newline dampinglike & $\mathcal{T}^{\rm even}_{\rm stagg}{\bf m}^{\rm A}\times(\hat{\bf p}\times{\bf m}^{\rm A})$ & $-\mathcal{T}^{\rm even}_{\rm stagg}{\bf m}^{\rm B}\times(\hat{\bf p}\times{\bf m}^{\rm B})$ \\ 
			\hline
			uniform\newline fieldlike & $\mathcal{T}^{\rm odd}_{\rm uni}{\bf m}^{\rm A}\times\hat{\bf p}$ & $-\mathcal{T}^{\rm odd}_{\rm uni}{\bf m}^{\rm B}\times\hat{\bf p}$ \\ 
			\hline			
			uniform\newline dampinglike & $\mathcal{T}^{\rm even}_{\rm uni}{\bf m}^{\rm A}\times(\hat{\bf p}\times{\bf m}^{\rm A})$ & $\mathcal{T}^{\rm even}_{\rm uni}{\bf m}^{\rm B}\times(\hat{\bf p}\times{\bf m}^{\rm B})$ \\ 
			\hline
			staggered\newline fieldlike & $\mathcal{T}^{\rm odd}_{\rm stagg}{\bf m}^{\rm A}\times\hat{\bf p}$ & $\mathcal{T}^{\rm odd}_{\rm stagg}{\bf m}^{\rm B}\times\hat{\bf p}$ \\ 		
			\hline
		\end{tabular}
		
		\caption{Table of four possible configurations of spin-orbit torques on two magnetic sublattices. Note that we assume staggered magnetization ${\bf m}^{\rm B}=-{\bf m}^{\rm A}$.}
		\label{table1}
	\end{table}

	Note that we assume the conventional lowest order form of spin-orbit torques and $\hat{\bf p}$ is the direction determined by the geometry, as we discuss in more detail in Sec.~\ref{Sec:CrI3 dynamics}. For now we note that the spin-orbit torque vanishes when $\hat{\mathbf{m}}^{\rm A,B}$ is aligned to $\hat{\bf p}$.  Symmetry dictates which spin-orbit torque terms are present.  For materials with inversion+time reversal symmetry, such as CuMnAs, the odd and even torques are uniform and staggered, respectively.  For materials with global inversion symmetry breaking, the odd and even torques are staggered and uniform, respectively~\cite{Zelezny2017}. These relations between the system symmetry and spin-orbit torque configuration apply when the magnetic sublattices are staggered. As the spin-orbit torque drives the spins out of the staggered configuration, symmetries are broken and the form of spin-orbit torque is no longer constrained to the forms given in Table~\ref{table1}. However we can still perform the standard stability analysis of Eq.~\ref{eq:LLG} at the fixed point where two magnetic sublattices are staggered and along the easy-axis.

	\begin{table}[htbp]
		\centering
		\begin{tabular}{ |c|c|c| } 
			\hline
			& staggered & uniform \\ 
			\hline
			even (dampinglike) & $\alpha\left(H_E+H_A\right)$  & $\alpha\sqrt{H_A\left(H_A+2H_E\right)}$  \\ 
			\hline
			odd (fieldlike) & $\sqrt{H_A\left(H_A+2H_E\right)}$ & $H_A$  \\ 
			\hline			
			
		\end{tabular}
		
		\caption{Expressions for the critical spin-orbit torque for different types of spin-orbit torque (dampinglike and fieldlike) and torque configurations (staggered and uniform).}
		\label{table2}
	\end{table}

	We consider each spin-orbit torque term individually and find the critical torque for inducing an instability.  The mathematical details are given in Appendix~\ref{app:stability} and the final results are summarized in Table~\ref{table2}.  The critical thresholds of time-reversal even torques include a small factor of $\alpha$, so that, provided $H_E$ is not too large, this torque may more easily excite N\'eel order dynamics. Identifying the potentially key role of the staggered dampinglike torque is a primary message of this work.  To illustrate an example in which this torque dominates the electrical excitation of an antiferromagnet, we next analyze the spin-orbit torque response of $n$-doped \ch{CrI3} in detail.\\

	\section{Spin dynamics in \ch{CrI3}: symmetry considerations}
	\label{Sec:CrI3 dynamics}
	
	In this section we analyze the constraints on the magnetic dynamics imposed by the symmetries of \ch{CrI3}.  The motivation for this is to reduce the degrees of freedom required to describe the system.  In general, a description of  antiferromagnets with moderate to weak exchange energy requires 4 degrees of freedom - an orientation $(\theta,\phi)$ for each spin.  This is in contrast to antiferromagnets in the large exchange limit, where the 2-dimensional N\'eel vector orientation is approximately sufficient to describe the system.  The 4-dimensional space of a general antiferromagnet is considerably more difficult to treat analytically.  However we will show that the symmetry of \ch{CrI3} enables a reduction of phase space to 2 dimensions.
	
	\begin{figure}[htbp]
		\includegraphics[width=0.9\columnwidth]{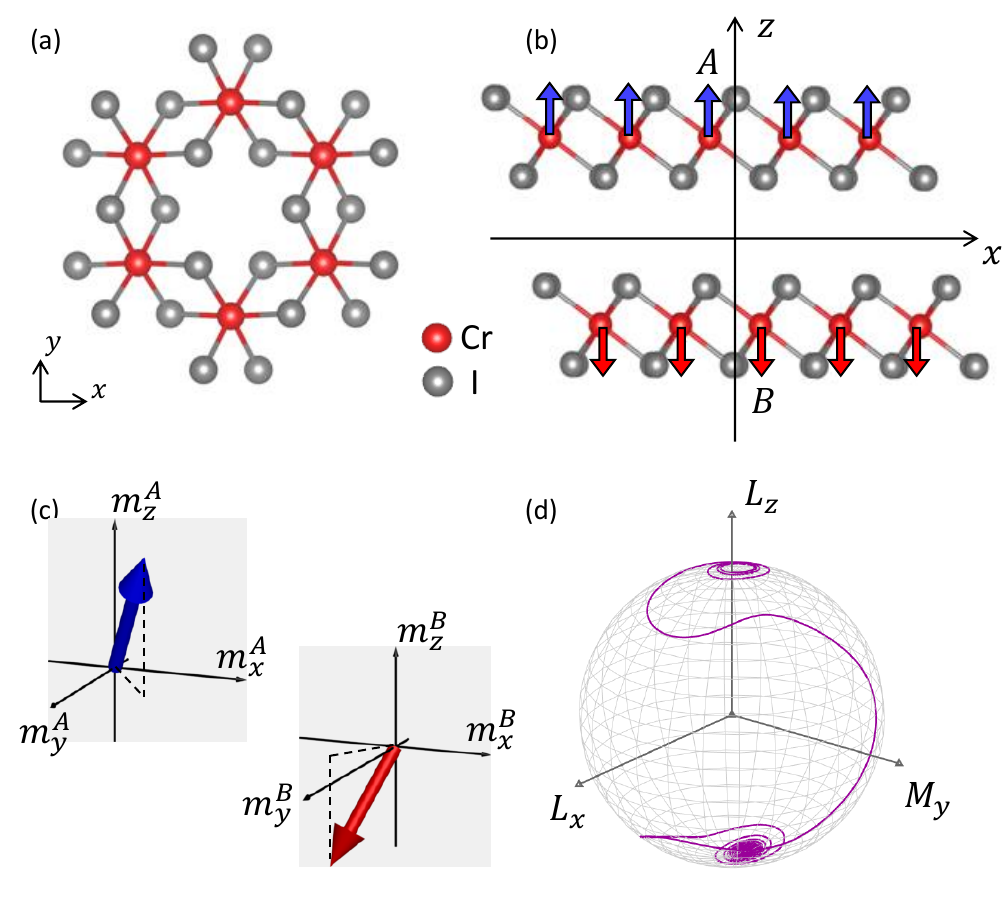}
		\caption{(Color online) (a) shows a top-down view of one layer of \ch{CrI3}.  The second layer (not shown) is displaced along the $x$-direction by a nearest-neighbor distance.  (b) Side view of \ch{CrI3}.  Note a lack of symmetry with respect to $x\rightarrow -x$. Other symmetries depend on the spin configuration: for a purely antiferromagnetic state, the system is invariant under inversion+time-reversal.  For a state with canting in the $y$ direction, the system has a 2-fold rotational symmetry about the $y$-axis. (c) Spin configurations on magnetic sublattices A and B considered in this work, with finite canting in the $y$-direction.  (d) Mixed representation of system spin in $\hat{{\bf N}}=(L_x,M_y,L_z)$ space, showing the spin-orbit torque switching trajectory of $\hat{{\bf N}}$ for applied electric field in the $y$-direction.
		}
		\label{Fig:coords}
	\end{figure}

	\ch{CrI3} is a recently discovered two-dimensional Van der Waals magnetic material~\cite{Huang2017,Gong2017,Deng2018}. There is intense recent interest in this material due to its unique and potentially useful properties, including the tunability and control of its magnetic state through gating and doping \cite{Jiang2018_NatureMaterials,Huang2018,Jiang2018,Fiete2020,Xiao2021} and its spin-filtering effects \cite{Song2018_TMR,Klein2018_TMR,Wang2018}. In the bilayer form of this semiconducting material, the two magnetic \ch{CrI3} layers are antiferromagnetically coupled and the ground state N\'eel vector is oriented perpendicular to the plane \cite{Huang2017} (see Fig.~\ref{Fig:coords}).  From the structure shown in Fig.~\ref{Fig:coords}(b), it's clear that bilayer \ch{CrI3} shares some characteristics with previously mentioned antiferromagnets, such as CuMnAs:  Inversion symmetry is locally broken on the magnetic sublattices (denoted A and B), while in the purely antiferromagnetic state, the bilayer is invariant under the combined operations of inversion plus time-reversal.  The exchange and anisotropy effective fields are similar in magnitude: $H_A\approx1.77~\rm T$, $H_E\approx0.76~\rm T$, and $\alpha\approx0.04$~\cite{Zhang2020}, so that, based on Table~\ref{table2}, the staggered dampinglike torque can be expected to play the dominant role in switching.
	
	We next discuss the symmetry of \ch{CrI3} which enables the degrees of freedom to be reduced from $(\hat{\mathbf{m}}^{\rm A},\hat{\mathbf{m}}^{\rm B})$ to a single orientation. We consider an applied electric field along the $y$-axis and spin configurations that are staggered in $(x,z)$ and uniform in $y$: $m^{\rm A}_{x/z}=-m^{\rm B}_{x/z}$ and $m^{\rm A}_y=m^{\rm B}_y$.  In this case, the system retains 2-fold rotation symmetry about the $y$-axis (see Fig.~\ref{Fig:coords}).  Any torque on the spins (including exchange and anisotropy) is therefore symmetry-constrained to satisfy $\mathcal{T}^{\rm A}_{x/z}=-\mathcal{T}^{\rm B}_{x/z}$ and $\mathcal{T}^{\rm A}_y=\mathcal{T}^{\rm B}_y$. The $x$ and $z$ components of the spins then remain staggered and the $y$ components remain uniform.  The trajectory of the spins is thus symmetry-confined to the subspace $(L_x,M_y,L_z)\equiv \frac{1}{2}(m_x^{\rm A}-m_x^{\rm B},~m_y^{\rm A}+m_y^{\rm B},~m_z^{\rm A}-m_z^{\rm B})$.  This motivates the definition of a ``mixed'' order parameter~\cite{footnote2}:
	\begin{eqnarray}
		{\bf N}\equiv \left(L_x,M_y,L_z\right)
	\end{eqnarray}
	Eq.~\ref{eq:LLG} leads to the following equation of motion for $\hat{{\bf N}}$:
	\begin{eqnarray}
		\label{eq:LLGn}
		\frac{d{\hat{\mathbf{N}}}}{dt}&=&	{\hat{\mathbf{N}}} \times \left(\frac{\gamma}{m}\frac{\delta E}{\delta {\hat{\mathbf{N}}}}  +\alpha\frac{d{\hat{\mathbf{N}}}}{dt} \right)+\mathcal{T}^{\rm odd}  \left( {\hat{\mathbf{N}}}\times\hat{\bf p}\right)\nonumber\\ &&~~~~+\mathcal{T}^{\rm even} ({\hat{\mathbf{N}}}\times (\hat{\bf p}\times{\hat{\mathbf{N}}})),
	\end{eqnarray}
	where the energy is comprised of the easy-axis anisotropy along $\hat{\bf z}$ and an effective hard-axis anisotropy along $\hat{\bf y}$, which encodes the magnetic exchange:
	\begin{eqnarray}
		E(\hat{\mathbf{N}})&=&-\frac{1}{2}mH_A\left(\hat{\mathbf{N}}\cdot \hat{\mathbf{z}}\right)^2 + m H_E\left(\hat{\mathbf{N}}\cdot \hat{\mathbf{y}}\right)^2.
	\end{eqnarray}
	We emphasize that the spin-orbit torque terms in Eq.~\ref{eq:LLGn} are staggered for the $x,z$ components and uniform for the $y$ component.  We compute these torques microscopically in the next section, where we find the angular dependence is more complex than the form given in Eq.~\ref{eq:LLGn}. Nevertheless the conclusions based on this simple form of spin-orbit torque are applicable to the results obtained with first principles calculations.

	We have verified that fluctuations away from the ${\bf N}$ subspace do not alter the steady state dynamics~\cite{footnote1}.  One important feature of this system which enables this simplification is that the easy-axis anisotropy is perpendicular to the axis of 2-fold rotation symmetry.  If this were not the case, then the ground state N\'eel vector would be aligned to the 2-fold rotation axis, and canting of the moments would destroy the 2-fold rotational symmetry.

	The simple form of the time evolution of $\hat{{\bf N}}$ allows for an intuitive description of the dynamics.	In the next section we show that $\hat{\bf p}$ has a standard $x$ component, and a $z$ component due to additional mirror plane symmetry breaking in \ch{CrI3}. For $\hat{\bf p}=\left(p_x,0,p_z\right)$, we again perform a stability analysis detailed in Appendix~\ref{app:Nspace}.  Note that this case differs slightly from the analysis presented earlier because $\hat{\bf p}$ is not aligned with the easy-axis, however the conclusion is the same.  The fixed points to lowest order in spin-orbit torque are $\hat{{\bf N}}=(-\frac{\mathcal{T}^{\rm odd} p_x}{\gamma H_A},\pm \frac{\mathcal{T}^{\rm even} p_x}{\gamma(2H_E+H_A)},\pm 1)$. The instability threshold to switch between fixed points is:
	\begin{equation}
		\label{eq:threshold}
		|\mathcal{T}^{\rm even} p_z|>\gamma\alpha(H_E+H_A).
	\end{equation}
	A typical switching trajectory is shown in Fig.~\ref{Fig:coords}(d): the spin-orbit torque drives $\hat{{\bf N}}$ from north pole to the fixed point close to south pole.\\

	\section{Microsopic calculations of spin-orbit torques in \ch{CrI3}} 
	\label{Sec:CrI3 SOT results}
	
	Having established the relevant degrees of freedom for the spin configuration in \ch{CrI3} as ${\bf N}$, we next present microscopic calculations of the spin-orbit torque per applied electric field - a quantity known as the ``torkance'' - as a function of $\hat{{\bf N}}$.  The procedure for this calculation is well-established \cite{Freimuth2014,Xue2020SOT}, and we briefly provide a description here and refer the reader to the Appendix~\ref{app:dft} for more technical details.  We first obtain the Hamiltonian in a localized atomic orbital basis using a combination of Quantum Espresso \cite{QEshort} and Wannier90 \cite{Wannier90}.  We then utilize linear response theory to compute the torkance on each magnetic sublattice.  We denote the $j^{\rm th}$ component of the torkance on atom $\rm A,B$ in response to an electric field along the $i$-direction with $\tau_{ij}^{\rm A,B}$.  The even and odd components of the torkance are given by:
	\begin{equation}
		\label{eq:eventorkance}
		\left(\tau^{\rm A,B}_{ij}\right)^{\rm even}=2e~\text{Im}\sum_{\substack{n,m\neq n}} f_{n} \frac{\left(\frac{\partial H}{\partial k_i}\right)_{n,m}\left(\mathcal{T}^{\rm A,B}_j\right)_{m,n}}{(E_{m}-E_{n})^2+\eta^2},
	\end{equation}
	\begin{equation}
		\label{eq:oddtorkance}
		\left(\tau^A_{ij}\right)^{\rm odd}=-e\sum_{\substack{n}}\frac{1}{2\eta}\frac{\partial f_{n}}{\partial E_n} \left(\frac{\partial H}{\partial k_i}\right)_{n,n}\left(\mathcal{T}^{\rm A,B}_j\right)_{n,n}.
	\end{equation}		
	The sum in Eqs.~\ref{eq:eventorkance}-\ref{eq:oddtorkance} is over eigenstates $|\psi_n\rangle$ of the ${\bf k}$-dependent Hamiltonian $H_{\bf k}$, where ${\bf k}$ is the Bloch wave vector and the eigenstate label $n$ includes ${\bf k}$ and band index.  $\left(O\right)_{n,m}=\langle \psi_n | O | \psi_m \rangle$ is the matrix element of the operator $O$, and $f_{n}=(e^{ (E_{n}-\mu)/k_{\rm B}T}+1)^{-1}$ is the equilibrium Fermi-Dirac distribution function. $\mu$ is the Fermi level, $\eta$ is the broadening parameter, and $e$ is the electron charge.  The atom-resolved torque operator is $\boldsymbol{\mathcal{T}}^{\rm A,B}=\frac{\rm i}{2\hbar}\left\{[\mathbf{S},\Delta],P^{\rm A,B}\right\}$, where $\mathbf{S}$ is the spin operator, $\Delta$ is the spin-dependent exchange-correlation potential, and $P^{\rm A(B)}$ is the projection operator onto the orbitals centered on atomic site A (B).  To compute the torque as a function of $\hat{{\bf N}}$, we manually rotate the spins on A and B sublattices.

	\begin{figure}[t]
		\includegraphics[width=1.\columnwidth]{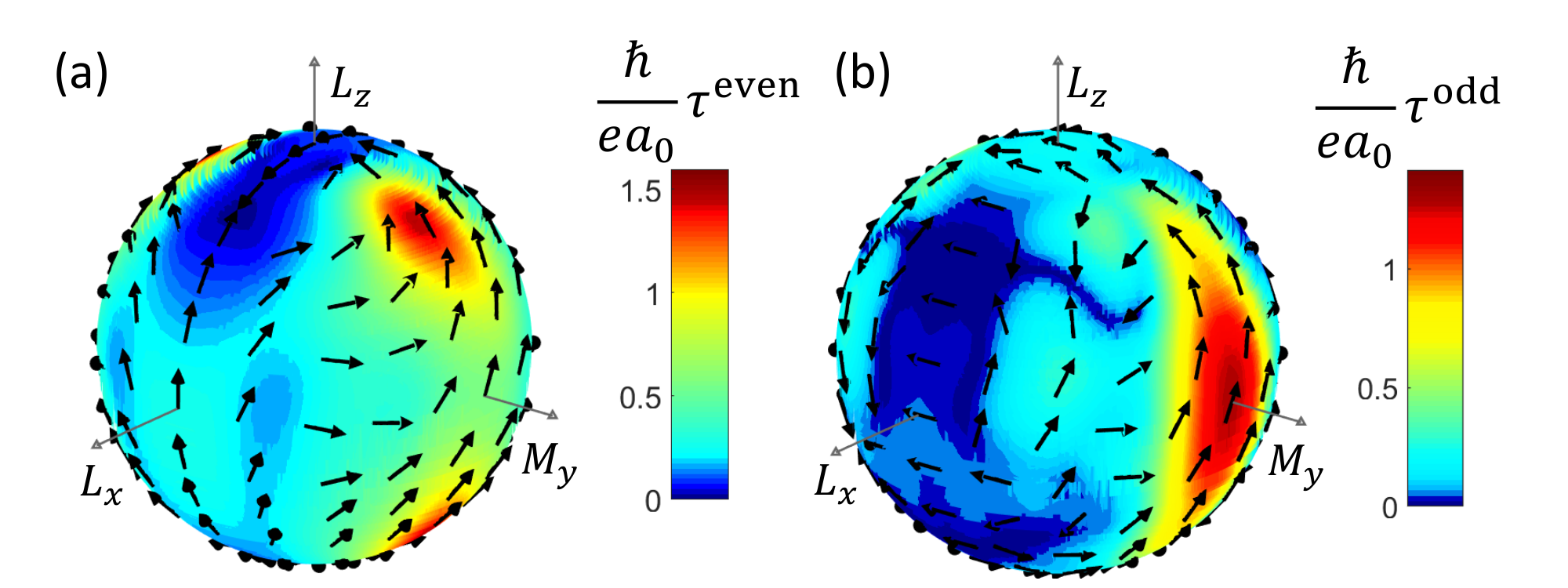}
		\caption{(Color online) Angular dependence of the dampinglike (a) and fieldlike (b) torkance on the $\hat{\mathbf{N}}$ direction $(\theta,\phi)$ for one layer of bilayer \ch{CrI3} under an external electric field along the $\hat{y}$ direction at Fermi level $\mu=50~\rm{meV}$ above the conduction band minimum. The arrow (color) on the sphere indicates the direction (magnitude) of the torkance at the given ${\bf N}$ direction. We use $k_BT=3~\rm{meV}, \eta=25~\rm{meV}$.
		}
		\label{Fig:GlobalTorque}
	\end{figure}	
	
	Figure~\ref{Fig:GlobalTorque} shows the $\hat{{\bf N}}$-dependence of the torkance with (a) and (b) showing the dampinglike (time-reversal even) and fieldlike (time-reversal odd) torkance, respectively.  The fixed points of both dampinglike and fieldlike torkance lie in the $L_x-L_z$ plane, away from the $L_z=0$ equator.  This is an important feature and is a consequence of the lack of mirror symmetry with respect to the $yz$ plane.  This position of the fixed point ensures that the spin-orbit torque drives $\hat{{\bf N}}$ to a point in the northern or southern hemisphere; after the spin-orbit torque is removed, $\hat{{\bf N}}$ then relaxes to the nearest easy-axis along $+\hat{\bf z}$ or $-\hat{\bf z}$. Previous studies on systems with similar in-plane mirror symmetry breaking, such as \ch{WTe2}-Py heterostructures~\cite{MacNeill2016,MacNeill2017,Stiehl2019,Xue2020SOT}, have verified that this symmetry breaking results in a spin-orbit torque that drives the magnetic order parameter to a point away from the equator.  Exploiting this property has emerged as an approach for deterministically switching perpendicularly magnetized thin films with spin-orbit torque, and we show here that this also enables switching of the perpendicular N\'eel vector.
	
	We note that the $\hat{{\bf N}}$-dependence of the torkance is quite complex, deviating substantially from the simple, lowest order form used in the analysis of the previous section.  In Appendix~\ref{app:sym}, we provide the full symmetry-allowed expansion of the torkance and quantify the substantial contribution from higher order terms.  We additionally find that the fixed points for even and odd torkance are different.   These features of the microscopically computed torkance have important consequences for the details of the dynamics of $\hat{{\bf N}}$ under spin-orbit torque, which we show in the next section.

	\begin{figure}[t]
		\includegraphics[width=1.\columnwidth]{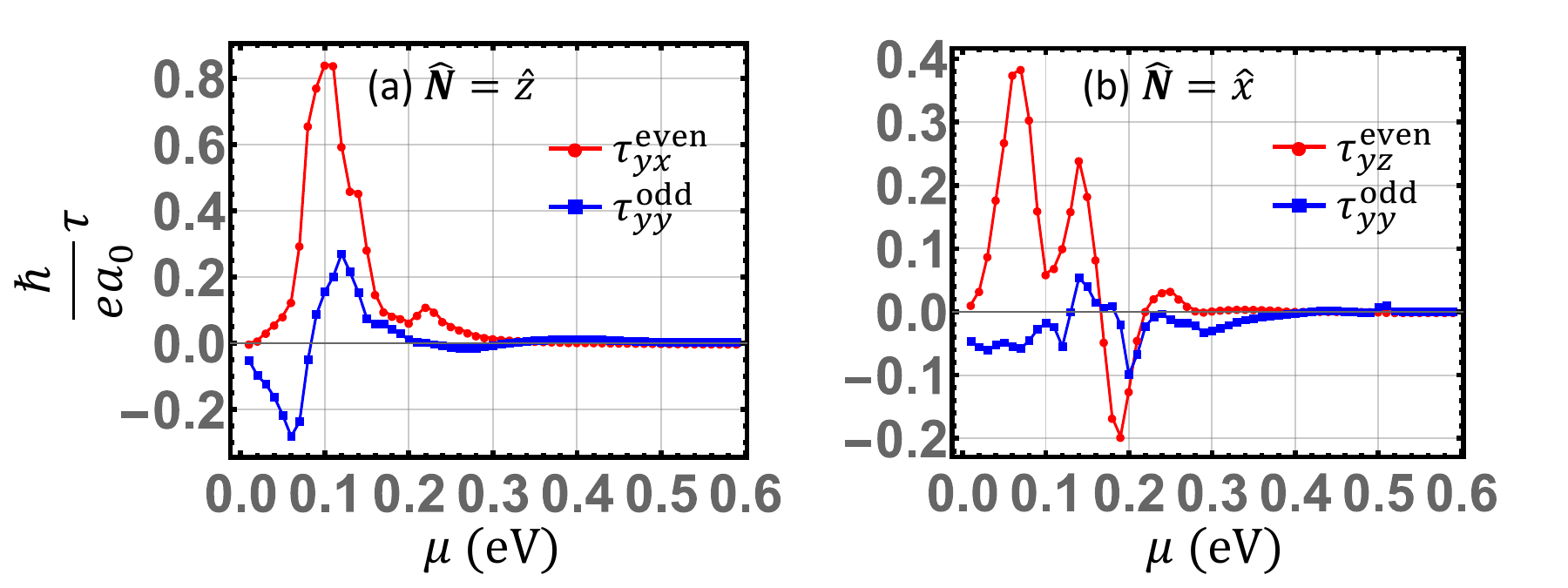}
		\caption{(Color online) Torkance as a function of chemical potential relative to the conduction band edge. The applied electric field is in $\hat{{\bf y}}$ direction. The $\hat{\mathbf{N}}$ vector is in $\hat{{\bf z}}$ (a) and $\hat{{\bf x}}$ (b). Red and blue lines represent staggered time-reversal even torkance and uniform time-reversal odd torque, respectively. The torkance for Fermi energies in the valence band are substantially smaller and not shown here.
		}
		\label{Fig2_mudependence}
	\end{figure}
	
	We next consider the torkance versus Fermi level for $\hat{{\bf N}}$ along $\hat{\bf z}$  and $\hat{{\bf x}}$ directions, shown in Figs.~\ref{Fig2_mudependence} (a) and (b), respectively.  Both even and odd components are peaked for Fermi energies near the conduction band minimum.  For $\hat{{\bf N}}=\hat{\bf z}$, the even torkance is approximately $1~ea_0/\hbar$ ($a_0\approx 0.0529~{\rm nm}$ is the Bohr radius) at $0.1~\rm{eV}$ above the conduction band minimum, which is larger than the even torkance in the ferromagnetic Pt/Co bilayer ($\approx0.6~ea_0/\hbar$) \cite{Freimuth2014}.  This large magnitude is due to band crossings in the conduction band from $p$-orbitals of the heavy Iodine atoms (Appendix~\ref{app:dft}). For $\hat{{\bf N}}=\hat{\bf x}$, the even torkance magnitude is around $0.4~ea_0/\hbar$. The even torkance for this $\hat{{\bf N}}$ configuration is solely a consequence of the in-plane mirror symmetry breaking.  This value is notably larger than the corresponding torkance derived from in-plane mirror symmetry breaking in the ferromagnetic 1T'-\ch{WTe2}/Co bilayer ($\approx0.1~ea_0/\hbar$) \cite{Xue2020SOT}. Note that the maximum torkance occurs at Fermi levels for which electrostatic doping might lead to a transition to a ferromagnetic ground state \cite{Huang2018,Jiang2018}. Nevertheless, appreciable spin-orbit torques are accessible at lower Fermi energies where antiferromagnetic order is retained.\\

	\begin{figure}[htbp]
		\includegraphics[width=1.\columnwidth]{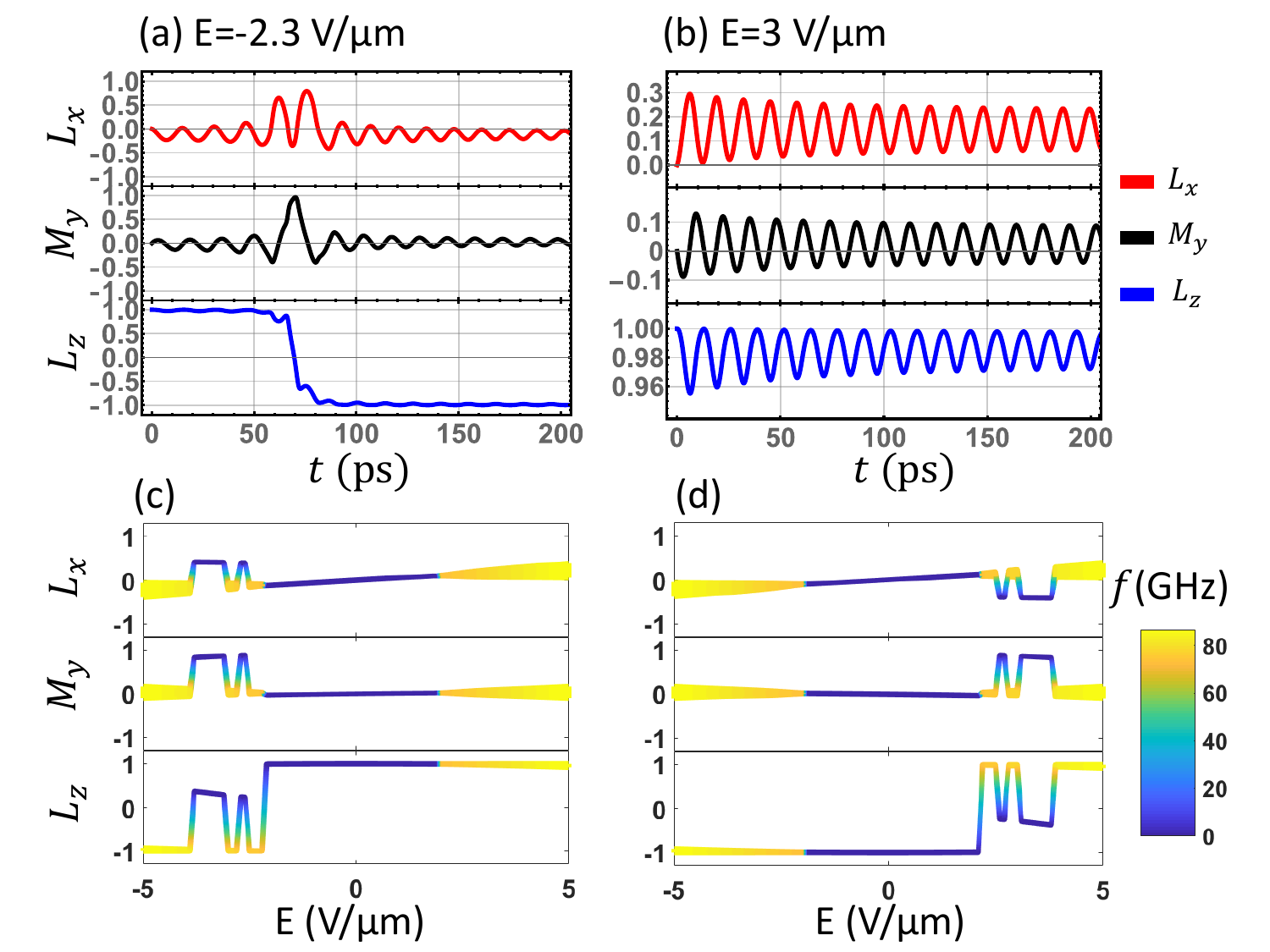}
		\caption{(Color online) Magnetization dynamics under spin-orbit torque, for applied electric field in the $\hat{\bf y}$ direction. (a) and (b) show the N\'eel and magnetization vector components as a function of time with applied electric field strength $-1.2$~V/{\textmu}m and $-3.5$~V/{\textmu}m, respectively. The initial configuration is $L_z=1$. Red, black, and blue lines represent the dynamics of $L_x$,$M_y$, and $L_z$ respectively.
			(c) and (d) show the final steady state of $\hat{\mathbf{N}}$ as a function of applied field with staring point at the $L_z=\pm1$ respectively. The spread in the $y$ coordinate indicates the oscillation amplitude, and the color of the spread represents the oscillation frequency.  
		}
		\label{Fig:LLG}
	\end{figure}
	
	\section{Spin dynamics in \ch{CrI3}: numerical results} 
	\label{Sec:CrI3 LLG results}

	Given the significant deviation of the $\hat{{\bf N}}$-dependence of the microscopically computed spin-orbit torque from the simple form of Eq.~\ref{eq:LLGn}, it's worthwhile to compute the spin dynamics with the \textit{ab initio} spin-orbit torque (Fig.~\ref{Fig:GlobalTorque}) as input into the coupled LLG equations (Eq.~\ref{eq:LLG}).  $\hat{\bf N}$ is parameterized by spherical coordinates $(\theta,\phi)$, and we use a bilinear interpolation of a dense $80\times 80$ mesh of spin-orbit torque obtained from first-principles to obtain the full $\hat{\bf N}$-dependence.
	
	Figure~\ref{Fig:LLG} shows the spin-orbit torque driven dynamics.  We find that the spin-orbit torque can either induce switching or induce steady state oscillations of $\hat{{\bf N}}$.  Figure~\ref{Fig:LLG}(a) shows that for an applied electric field $E={\rm -2.3~V/\mu m}$, the spin-orbit torque switches the N\'eel order $L_z$ from the north pole to the southern hemisphere within 100~ps and generates a finite in-plane magnetization $M_y$. Note that the input spin-orbit torque terms include both dampinglike  and fieldlike torques (Fig.~\ref{Fig:GlobalTorque}). However, dampinglike torque only has to compete with the product of anisotropy plus exchange and the small damping factor (Table~\ref{table2}). We expect dampinglike would play a more important role when the magnitudes of dampinglike and fieldlike torque
	are comparable. Indeed, by separately turning off the fieldlike (odd) or dampinglike (even) contributions to the spin-orbit torque, we find that the switching of $L_z$ originates from the dampinglike torque, while the fieldlike torque helps to accelerate the switching dynamics and reduce the switching E-field threshold.
	Figure~\ref{Fig:LLG}(b) shows an oscillating steady state for $E={\rm 3~V/\mu m}$, with a frequency of approximately 80~GHz.  We find that both dampinglike and fieldlike torque are required to induce steady state oscillation.

	We summarize the final steady states as a function of the applied field $E$ for two initial magnetization configurations $L_z=+1$ and $L_z=-1$ in Fig.~\ref{Fig:LLG} (c) and (d), respectively.  The switching of the N\'eel vector occurs at approximately $|E|=2~{\rm V/\mu m}$.  This threshold compares well with the estimate provided by Eq.~\ref{eq:threshold}.  Reaching the larger scale oscillations at large applied $E$ will rely on the material to sustain large power dissipation, which depends in turn on factors such as the carrier mobility.  The flatness of the conduction bands implies a low mobility, as seen experimentally \cite{patil2020intriguing}, which should enable larger applied electric fields.  Fig.~\ref{Fig:LLG} (c) and (d) demonstrate hysteretic switching of the N\'eel vector, and are related by mirror symmetry about the $xz$ plane.

\begin{figure}[htbp]
		\includegraphics[width=1\columnwidth]{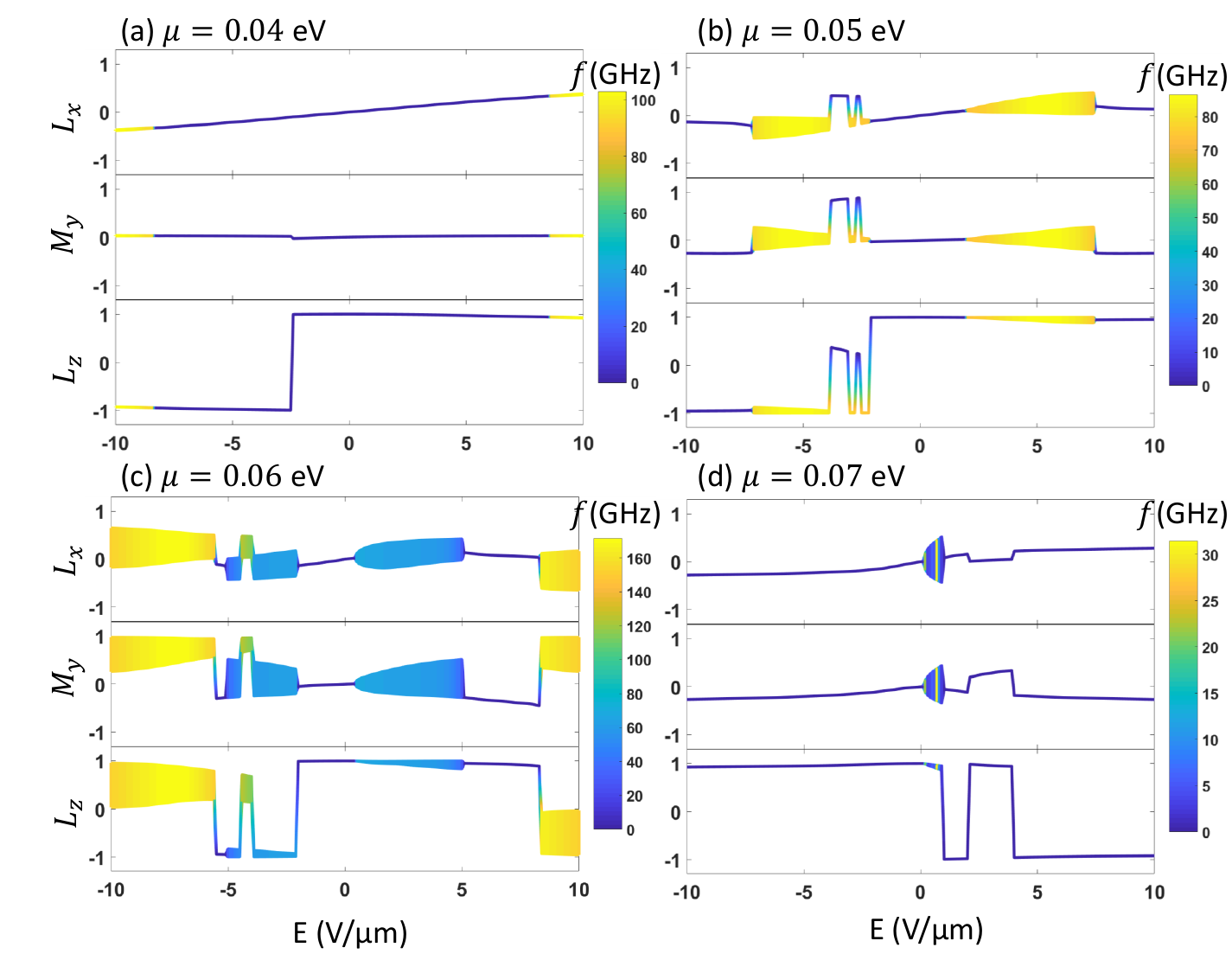}
		
		\caption{(Color online) Final steady state of $\hat{\mathbf{N}}$ as a function of applied field with staring point at the $L_z=+1$ for various chemical potentials respectively. The spread in the $y$ coordinate indicates the oscillation amplitude, and the color of the spread represents the oscillation frequency.
		}
		\label{fig:mu}
	\end{figure}

	Before we conclude, we include additional plots of final steady states at different Fermi levels summarized in Fig.~\ref{fig:mu}. Both switching and oscillating behaviors can be observed at various chemical potentials and electric-field strengths.The chemical potential can be tuned by perpendicular gate voltage in principle and Fig.~\ref{fig:mu} indicates bilayer \ch{CrI3} can have tunable functions by controlling both in-plane and out-of-plane fields. Recent experiments~\cite{Jiang2018} also demonstrate a magnetic phase transition from an antiferromagnet ground state to a ferromagnetic state under electron doping. We ignore this transition to compute the torque at higher electron densities in the antiferromagnetic state, for the sake of gaining an understanding of how this electronic structure influences the torque.  A comprehensive study of the ground state transitions along with their spin-orbit torque responses is beyond the scope of the current work.
	\\

	\section{Discussion} 
	\label{Sec:Discussion}
	The experimental detection of the N\'eel vector reversal is challenging. For bilayer \ch{CrI3}, out-of-plane magnetic-optical Kerr effect (MOKE) imaging has previously been used to discriminate between $L_z=+1$ and $L_z=-1$~\cite{Huang2018}, and transport effects such as nonlinear anisotropic magnetoresistance can also detect ${\bf N}$~\cite{Godinho2018}.  We also note that the moderate exchange energy leads to the development of a substantial steady state in-plane magnetization of the driven system, which may be detected experimentally with in-plane MOKE. 
	
	Aside from the particulars of \ch{CrI3}, in this work we show generally that antiferromagnets in the weak to moderate exchange coupling regime exhibit different behaviors from their more commonly studied large $H_E$ counterparts.  The switching criteria for these antiferromagnets is reduced by a factor of magnetic damping, offering potentially easier routes to electrical manipulation.  Continued progress in the field of Van der Waals antiferromagnets should provide further opportunities for unique modes of electrical control of these materials.

	\section{Acknowledgement}
	F.X. acknowledges support under the Cooperative Research Agreement between the University of Maryland and the National Institute of Standards and Technology Physical Measurement Laboratory, Award 70NANB14H209, through the University of Maryland.
	
	\bibliography{reference}{}

\begin{thebibliography}{54}
\bibitem [{foo({\natexlab{b}})}]{footnote2}%
  \BibitemOpen
  \bibinfo {note} {For spin configurations in the ${\bf N}$-subspace
  ($m_{x/z}^{\rm A}=-m_{x/z}^{\rm B},~m_y^{\rm A}=m_y^{\rm B}$), it's easy to
  show that $|{\bf N}|=|{\hat {\bf m}}^{\rm A}|=1$. This spin configuration
  subspace can also be represented with ${\hat {\bf m}}^{\rm A}$ alone, with
  the value of ${\hat {\bf m}}^{\rm B}$ implied by symmetry.}\BibitemShut
  {Stop}%
\bibitem [{foo({\natexlab{c}})}]{footnote1}%
  \BibitemOpen
  \bibinfo {note} {To determine the robustness of the projection of the spin
  configuration to the ${\bf N}$ subspace, we solved for the steady state for
  $20^4$ possible initial conditions in the 4-dimensional space $(\theta^{\rm
  A},\phi^{\rm A},\theta^{\rm B},\phi^{\rm B})$, which characterize the
  orientations of the spins on A and B. We find that the steady states from all
  initial conditions belong to the ${\bf N}$ subspace.}\BibitemShut {Stop}%
\end{thebibliography}
	\bibliographystyle{apsrev4-2}
	
	\appendix
	\section{General stability analysis}	
	\label{app:stability}
	The dynamics of two coupled spins A and B are described the set of Landau-Lifshitz-Gilbert (LLG) equations: \cite{Stiles2006,Gomonay2010,Manchon2017}:
	\begin{widetext}
		\begin{eqnarray}
			\label{eq:LLG_app}
			\frac{d\hat{\mathbf{m}}^{\rm A}}{dt}-\alpha\hat{\mathbf{m}}^{\rm A}\times\frac{d\hat{\mathbf{m}}^{\rm A}}{dt}&=&-\gamma H_A \left(\hat{\mathbf{m}}^{\rm A}\times 	\hat{{\bf z}}\right)\left(\hat{\mathbf{m}}^{\rm A}\cdot\hat{{\bf z}}\right)+\gamma H_E\left(\hat{\mathbf{m}}^{\rm A}\times\hat{\mathbf{m}}^{\rm B}\right)+\boldsymbol{\mathcal{T}}^{\rm A},\nonumber\\
			\frac{d\hat{\mathbf{m}}^{\rm B}}{dt}-\alpha\hat{\mathbf{m}}^{\rm B}\times\frac{d\hat{\mathbf{m}}^{\rm B}}{dt}&=&-\gamma H_A \left(\hat{\mathbf{m}}^{\rm B}\times 	\hat{{\bf z}}\right)\left(\hat{\mathbf{m}}^{\rm B}\cdot\hat{{\bf z}}\right)+\gamma H_E\left(\hat{\mathbf{m}}^{\rm B}\times\hat{\mathbf{m}}^{\rm A}\right)+\boldsymbol{\mathcal{T}}^{\rm B}
		\end{eqnarray}	
	\end{widetext}
	where $\gamma$ is the absolute value of the electron gyromagnetic ratio, $H_A$ is the magnetic anisotropy field strength, $\hat{{\bf z}}$ is the magnetic easy-axis, $H_E$ is the antiferromagnetic exchange field, $\alpha$ is the damping parameter, and  $\boldsymbol{\mathcal{T}}^{\rm (A,B)}$ is the spin-orbit torque on the ${\rm (A,B)}$ sublattice.  It's convenient to work in spherical coordinates, where the magnetization vector is given by $\hat{\mathbf{m}}=\left(\sin\theta \cos\phi,\sin\theta \sin\phi,\cos\theta\right)$.  The torque is always perpendicular to the magnetization, so that it can be expressed in terms of the $\mathbf{e}_\theta,\mathbf{e}_\phi$ components, where $\mathbf{e}_\theta\equiv(\cos\theta\cos\phi,\cos\theta\sin\phi,-\sin\theta)$ and $\mathbf{e}_\phi\equiv(-\sin\phi,\cos\phi,0)$. The matrix form of Eq.~\ref{eq:LLG} is
	\begin{equation}
		\label{eq:LLG2}
		\begin{pmatrix}
			\dot\phi^{\rm A} \\
			\dot\theta^{\rm A} \\
			\dot\phi^{\rm B}\\
			\dot\theta^{\rm B}
		\end{pmatrix}=\frac{1}{1+\alpha^2}
		\begin{pmatrix}
			\frac{1}{\sin\theta^{\rm A}}&\frac{\alpha}{\sin\theta^{\rm A}}&0&0\\
			-\alpha&1&0&0\\
			0&0&\frac{1}{\sin\theta^{\rm B}}&\frac{\alpha}{\sin\theta^{\rm B}}\\
			0&0&-\alpha&1
		\end{pmatrix}
		\begin{pmatrix}
			T^{\rm A}_\phi\\
			T^{\rm A}_\theta\\
			T^{\rm B}_\phi\\
			T^{\rm B}_\theta
		\end{pmatrix},
	\end{equation}
	where $T^{\phi,\theta}_{A,B}$ is obtained by projecting the right hand side of Eq.~\ref{eq:LLG} to the $\mathbf{e}_{\phi,\theta}$ directions on the A and B sublattices.  
	
	The fixed points and their stability are determined by the set of torque expressions $\Gamma=\left(T^{\rm A}_\phi,T^{\rm A}_\theta,T^{\rm B}_\phi,T^{\rm B}_\phi\right)$.  A fixed points satisfies $\Gamma=0$, and its stability is determined by the eigenvalues of the dynamic matrix $D$ \cite{Bazaliy2004}.  $D$ is given by the product of the matrix given on the right-hand-side of Eq.~\ref{eq:LLG2} and the Jacobian matrix derived from $\Gamma$ evaluated at the fixed point.  A fixed point goes from stable to unstable as the real part of its eigenvalue goes from negative to positive.

	We consider how collinear antiferromagnets become unstable against different types of spin-orbit torques. We can decompose the current-induced spin-orbit torques to four distinct contributions depending on the time-reversal symmetry and whether the torques on two sublattices are uniform or opposite shown in the Table~\ref{table1}. We assume the conventional lowest order form of spin-orbit torque, as shown in Table \ref{table1}.  As discussed in the main text, the direction $\hat{\bf p}$ is determined by the system symmetry. Depending on the relative sign of constant prefactor $\mathcal{T}^{\rm even,odd}$ on two sublattices, spin-orbit torques on two sublattices are either uniform or staggered.  
	

	In the following analysis, we take the easy-axis to be $\bf y$ and $\mathbf{p}=\mathbf{y}$. This is for the convenience of avoiding the singular spherical coordinates near the north and south poles. The fixed points we evaluate are $L_y=\pm 1$, or $\theta^{\rm A}=\theta^{\rm B}=\pi/2$, $\phi^{\rm A}=-\phi^{\rm B}=\pm\pi/2$. Note that it is necessary to evaluate the full $4\times 4$ Jacobian matrix derived from Eq.~\ref{eq:LLG2}.  We consider the four different configurations of spin-orbit torque (even/odd, staggered/uniform) individually below.\\

	1.  {\it Staggered dampinglike torque}: The dynamic matrix $D$ up to the linear order of Gilbert damping $\alpha$ at $L_y=+1$ is 
	\begin{widetext}
		\begin{equation}
			D=\begin{pmatrix}
				-\alpha(H_E+H_A)-\mathcal{T}^{\rm even} &H_E+H_A-\alpha \mathcal{T}^{\rm even}&\alpha H_E& H_E\\
				-(H_E+H_A)+\alpha \mathcal{T}^{\rm even}&-\alpha(H_E+H_A)-\mathcal{T}^{\rm even}& H_E&-\alpha H_E\\
				\alpha H_E& H_E&-\alpha(H_E+H_A)-\mathcal{T}^{\rm even} &H_E+H_A-\alpha \mathcal{T}^{\rm even}\\
				H_E&-\alpha H_E&-(H_E+H_A)+\alpha \mathcal{T}^{\rm even}&-\alpha(H_E+H_A)-\mathcal{T}^{\rm even} 
			\end{pmatrix}.
		\end{equation}
		
		We obtain the eigenvalues of the dynamic matrix $D$ as:
		\begin{eqnarray}
			\lambda=-\mathcal{T}^{\rm even}-\alpha(H_E+H_A) \pm \sqrt{-2H_E H_A-H_A^2+2(H_E+H_A)\mathcal{T}^{\rm even}\alpha+H_E^2\alpha^2-\mathcal{T}^{\rm{even}2}\alpha^2},
		\end{eqnarray}
	\end{widetext}
	The square root is an imaginary number since $H_A, H_E$ are positive numbers. The real part of $\lambda$ becomes positive when $\mathcal{T}^{\rm even}<-\alpha(H_E+H_A)$. 
	This instability threshold has the advantage that a small damping factor can help reduce the required electrical field or current. However it is difficult to achieve in the limit of very large exchange coupling strength. \\

	2.  {\it Uniform fieldlike torque}: The dynamic matrix evaluated at $L_y=1$ is:
	\begin{widetext}
		\begin{equation}
			D=\begin{pmatrix}
				-\alpha(H_E+H_A-\mathcal{T}^{\rm odd}) &H_E+H_A-\mathcal{T}^{\rm odd}&\alpha H_E& H_E\\
				-(H_E+H_A)+\mathcal{T}^{\rm odd}&-\alpha(H_E+H_A-\mathcal{T}^{\rm odd}) & H_E&-\alpha H_E\\
				\alpha H_E& H_E&-\alpha(H_E+H_A-\mathcal{T}^{\rm odd})  &H_E+H_A-\mathcal{T}^{\rm odd}\\
				H_E&-\alpha H_E&-(H_E+H_A)+ \mathcal{T}^{\rm odd}&-\alpha(H_E+H_A-\mathcal{T}^{\rm odd})  
			\end{pmatrix}.
		\end{equation}	
		The eigenvalues of the resulting dynamic matrix are:
		\begin{eqnarray}
			\lambda=-\alpha(H_E+H_A-\mathcal{T}^{\rm odd}) \pm \sqrt{H_E^2+H_E^2\alpha^2-(H_E+H_A-\mathcal{T}^{\rm odd})^2}.
		\end{eqnarray}	
	\end{widetext}
	
	Note that both eigenvalues are doubly degenerate. The condition of having positive real part of one eigenvalue (with positive square root) is $\mathcal{T}^{\rm odd}>H_A$. In other words, $\mathcal{T}^{\rm odd}$ is competing with the easy-axis anisotropy by reducing the effective anisotropy field. This mechanism has the advantage that a large exchange field does not affect its effectiveness. \\

	3. {\it Uniform dampinglike torque}: The dynamic matrix evaluated at $L_y=1$ is:
	\begin{widetext}
		\begin{equation}
			D=\begin{pmatrix}
				-\alpha(H_E+H_A)-\mathcal{T}^{\rm even} &H_E+H_A-\alpha \mathcal{T}^{\rm even}&\alpha H_E& H_E\\
				-(H_E+H_A)+\alpha \mathcal{T}^{\rm even}&-\alpha(H_E+H_A)-\mathcal{T}^{\rm even}& H_E&-\alpha H_E\\
				\alpha H_E& H_E&-\alpha(H_E+H_A)+\mathcal{T}^{\rm even} &H_E+H_A+\alpha \mathcal{T}^{\rm even}\\
				H_E&-\alpha H_E&-(H_E+H_A)-\alpha \mathcal{T}^{\rm even}&-\alpha(H_E+H_A)+\mathcal{T}^{\rm even} 
			\end{pmatrix}.
		\end{equation}
		This matrix does not have analytic solutions. However, we can obtain approximate eigenvalues by assuming $\mathcal{T}^{\rm even}$ is proportional to $\alpha$ and then dropping terms $\alpha \mathcal{T}^{\rm even}$ (because we expand every term up to 1st order of small damping factor $\alpha$). Then analytic eigenvalues up to the first order of $\alpha$ are:	
		\begin{eqnarray}
			\lambda=-\alpha(H_E+H_A) \pm \sqrt{\pm2 i|\mathcal{T}^{\rm even}|(H_E+H_A)-2H_E H_A-H_A^2+\mathcal{T}^{\rm{even}2}}.
		\end{eqnarray}
	\end{widetext}
	The condition of having positive real part of eigenvalues up to the 1st order of $\alpha$ is then $|\mathcal{T}^{\rm{even}}|>\alpha\sqrt{H_A(H_A+2H_E)}$. This threshold has the advantage of reducing threshold by the damping factor. The realization of this torque requires global inversion symmetry breaking, such as found in heterostructures composed of antiferromagnets and heavy metals.\\

	4. {\it Staggered fieldlike torque}: The dynamic matrix evaluated at $L_y=1$ is:
	\begin{widetext}
		\begin{equation}
			D=\begin{pmatrix}
				-\alpha(H_E+H_A-\mathcal{T}^{\rm odd}) &H_E+H_A-\mathcal{T}^{\rm odd}&\alpha H_E& H_E\\
				-(H_E+H_A)+\mathcal{T}^{\rm odd}&-\alpha(H_E+H_A-\mathcal{T}^{\rm odd}) & H_E&-\alpha H_E\\
				\alpha H_E& H_E&-\alpha(H_E+H_A+\mathcal{T}^{\rm odd})  &H_E+H_A+\mathcal{T}^{\rm odd}\\
				H_E&-\alpha H_E&-(H_E+H_A+ \mathcal{T}^{\rm odd})&-\alpha(H_E+H_A+\mathcal{T}^{\rm odd})  
			\end{pmatrix}.
		\end{equation}	
	\end{widetext}	
	The analytic eigenvalue solutions are not available for this case. However, we numerically find the instability threshold to be $|\mathcal{T}^{\rm odd}|>\sqrt{H_A(H_A+2H_E)}$.\\
	
	These four cases are the simplest four ways to manipulate and control the antiferromagnetic order with spin-orbit torques. Depending on the relative parameters, different mechanisms can be favored to switch the order N\'eel vector. For example, uniform fieldlike torque or uniform dampinglike torque can be favored when $H_E \gg H_A$ while staggered dampinglike torque can be favored when $\alpha(H_A+H_E)$ is much smaller than other thresholds.\\

	\section{Analysis of the ${\bf N}$ subspace}
	\label{app:Nspace}
	As discussed in the main text, the 2-fold rotational symmetry about the $y$-direction constrains the spins to the subspace spanned by ${\bf N}=(L_x,M_y,L_z)$.  Due to the lack of mirror symmetry about the $yz$ plane, the lowest order fieldlike and dampinglike torque have the form of $\mathcal{T}^{\rm odd}\hat{\mathbf{m}}\times\mathbf{p}$ and $\mathcal{T}^{\rm even}\hat{\mathbf{m}}\times(\mathbf{p}\times\hat{\mathbf{m}})$, respectively, where $\mathbf{p}=(p_x,0,p_z)$. 2-fold rotational symmetry about the $y$-axis leads to the following relation between any torque on A and B sublattices:
	\begin{eqnarray}
		{T}_y^{\rm A}&=&{T}_y^{\rm B},\\
		{T}_{x,z}^{\rm A}&=&-{T}_{x,z}^{\rm B},	
	\end{eqnarray}
	where torque $T$ includes every term on the right hand side of Eq.~\ref{eq:LLG}, i.e., anisotropy, exchange, and spin-orbit torque.
	The anisotropy field gives rise to the stable initial state $(L_x,M_y,L_z)=(0,0,\pm1)$ and we are interested in the condition where the spin-orbit torque drives the system away from the equilibrium state. 
	To avoid the singular spherical coordinates near these points, we perform an index permutation $(x,y,z) \rightarrow (z,x,y)$, so that the magnetic subspace is now labelled by $\left(M_x,L_y,L_z\right)$.

	In the subspace of $(M_x,L_y,L_z)$, $\theta^{\rm B}=\pi-\theta^{\rm A}, \phi^{\rm B}=-\phi^{\rm A}$. We can verify that the torque in Eq.~\ref{eq:LLG2} are staggered, so the $4\times4$ matrix form of LLG equation becomes two identical $2\times2$ matrices:
	\begin{eqnarray}
		\label{eq:LLG_Nspace}
		\begin{pmatrix}
			\dot\phi^{\rm A} \\
			\dot\theta^{\rm A}
		\end{pmatrix}&=&\frac{1}{1+\alpha^2}
		\begin{pmatrix}
			\frac{1}{\sin\theta^{\rm A}}&\frac{\alpha}{\sin\theta^{\rm A}}\\
			-\alpha&1
		\end{pmatrix}
		\begin{pmatrix}
			T^{\rm A}_\phi\\
			T^{\rm A}_\theta\\
		\end{pmatrix},\\
		\begin{pmatrix}
			-\dot\phi^{\rm A} \\
			-\dot\theta^{\rm A}
		\end{pmatrix}&=&\frac{1}{1+\alpha^2}
		\begin{pmatrix}
			\frac{1}{\sin\theta^{\rm A}}&\frac{\alpha}{\sin\theta^{\rm A}}\\
			-\alpha&1
		\end{pmatrix}
		\begin{pmatrix}
			-T^{\rm A}_\phi\\
			-T^{\rm A}_\theta\\
		\end{pmatrix}.
	\end{eqnarray}	
	Now we can drop the sublattice subscript and the equilibrium state is obtained by solving the equations $(T_{\phi},T_{\theta})=0$. We can find solutions to this set of nonlinear equations with the ansatz $\theta=\pi/2+a,\phi=\pi/2+b$ where $a,b\ll 1$ by assuming small spin-orbit torque terms. By expanding all terms up to the first order of spin-orbit torques, we find:
	\begin{eqnarray}
		a=\frac{\mathcal{T}^{\rm odd} p_z }{\gamma H_A}, ~~b=-\frac{\mathcal{T}^{\rm even} p_z }{\gamma(2H_E+H_A)}.
	\end{eqnarray}
	This equilibrium corresponds to the magnetization configuration $(M_x,L_y,L_z)=\left(\frac{\mathcal{T}^{\rm even} p_z }{\gamma\left(2H_E+H_A\right)},1,-\frac{\mathcal{T}^{\rm odd} p_z}{H_A}\right)$. 
	
	The dynamic matrix $D$ up to the linear order of $\alpha,a,b,$ and the spin-orbit torque terms is
	\begin{equation}
		D=\begin{pmatrix}
			-\mathcal{T}^{\rm even} p_y-\alpha(2H_E+H_A) & H_A-\mathcal{T}^{\rm odd} p_y \\
			-2H_E-H_A+\mathcal{T}^{\rm odd}p_y  & -\mathcal{T}^{\rm even} p_y-H_A\alpha
		\end{pmatrix}.
	\end{equation}
	The two eigenvalues are
	\begin{eqnarray}
		\lambda&=&-\mathcal{T}^{\rm even}p_y-(H_E+H_A)\alpha\nonumber \\ &&~\pm i\left[\sqrt{H_A(2H_E+H_A)}-\frac{\mathcal{T}^{\rm odd}p_y H_E}{\sqrt{H_A(2H_E+H_A)}}\right].
	\end{eqnarray}
	The switching condition is then $p_y\mathcal{T}^{\rm even}<-\alpha(H_E+H_A)$. This analysis reveals the key ingredients of staggered dampinglike torque: the torque component along the direction perpendicular to the easy-axis drives the net magnetization along the direction perpendicular to both torque direction and easy-axis direction while the torque component along the direction parallel to the easy-axis switches the N\'eel order from one hemisphere to the other. Comparing to the fieldlike torque, the dampinglike torque only needs to compete with the total strength of exchange and anisotropy field multiplying a small Gilbert damping factor. The staggered dampinglike torque is therefore more favored to drive the AFM system when the exchange and anisotropy field have the same order of magnitude.
	
	\section{First-principles details}
	\label{app:dft}
	\begin{figure}[htbp]
		\includegraphics[width=1\columnwidth]{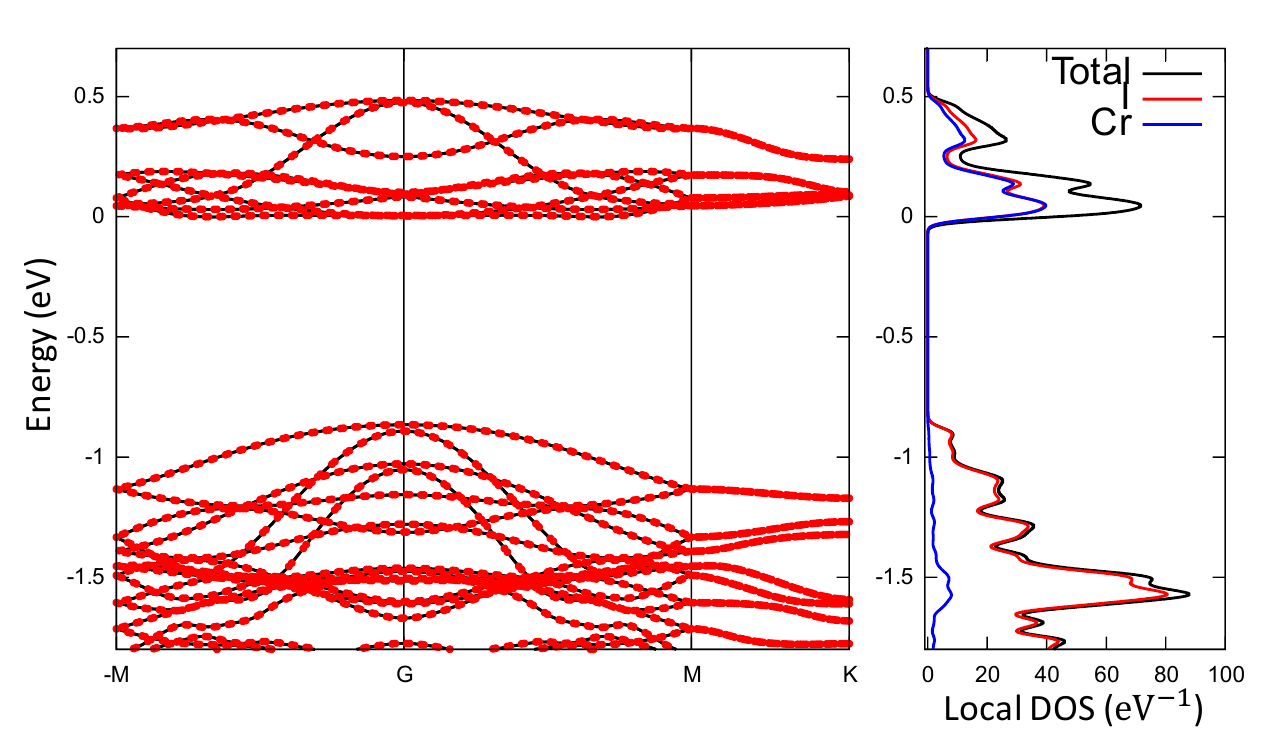}
		\caption{(Color online) Bandstructure of bilayer \ch{CrI3} along the -M$(-\frac{1}{2},0,0)$, G$(0,0,0)$, M$(\frac{1}{2},0,0)$, K$(\frac{1}{2},\frac{1}{2},0)$ line in ${\bf k}$-space and the projected density of states. In the bandstructure, red dots represent bands obtained from final tight-binding Hamiltonian while black lines represent the bands obtained from plane-wave basis.
			The red line, blue line and the black line represent the Iodine atoms, Cr atoms, and total atoms contribution to the local density of states, respectively. Note that up spins and down spins are degenerate because of the $PT$ symmetry and we do not include spin-orbit coupling in these plots.
		}
		\label{Fig:bands}
	\end{figure}		
	We use Quantum ESPRESSO \cite{QE} to compute the electronic structure of bilayer \ch{CrI3}. We adopt the experimental unit cell parameters \cite{Zhang2020} of bilayer \ch{CrI3} (space group C2/m): $a=0.6904~\text{nm},~b=1.1899~\text{nm},~c=0.7008~\text{nm}$, and $\beta=108.74 $\textdegree. In the Quantum ESPRESSO implementation, we use the pseudopotentials from PSlibrary \cite{DALCORSO2014337} generated with a scalar relativistic calculation using Projector Augmented-Wave method \cite{PAW} and Perdew-Burke-Ernzerhof exchange correlations \cite{PBEGGA}. We utilize a $7\times 12 \times 1$ Monkhorst-Pack mesh \cite{MPmesh}, $1360~{\rm eV}$ cutoff energy, $1.36\times10^{-3}~{\rm eV}$ total energy convergence threshold , and $0.08~{\rm eV/nm}$ force convergence threshold. We add a Hubbard on-site energy $U=3~\rm eV$ on Cr atoms \cite{Sivadas2018}. We next utilize Wannier90 \cite{Wannier90} to obtain the Hamiltonian in an atomic basis.  We project plane-wave solutions onto atomic $s,d$ orbitals of Cr atoms, $p$ orbitals of I atoms. We then symmetrize the Wannier-like tight-binding Hamiltonian using TBmodels \cite{TBmodels} since the presence of slight asymmetry in the tight-binding Hamiltonian results in symmetry-disallowed torque, and we remove small spin-dependent hopping terms. The final symmetrized tight-binding band structures match those obtained with plane-wave methods.  We add on-site spin-orbit coupling terms $\alpha \mathbf{L}\cdot\mathbf{S}$, where ${\bf L}$ and ${\bf S}$ are the orbital angular momentum and spin operators, respectively.  We use $\alpha=\left[90,~580\right]~{\rm meV}$ for Cr, and I \cite{Tartagliaeabb9379}.  Adding spin-orbit coupling ``by hand'' in this manner requires that Wannier orbitals are not localized in order to ensure their forms are spherical harmonics consistent with the standard representation of ${\mathbf{L}}$. We adopt this approach because it is technically easier to achieve a good Wannier projection of a collinear magnetized Hamiltonian, and the on-site spin-orbit coupling approximation yields accurate results (see Fig.~\ref{Fig:bands} to see a comparison of band structure obtained with Quantum ESPRESSO and Wannier orbitals). We use a dense ${\bf k}$ mesh of $400\times232$ to evaluate the torkance, given by Eqs.~5 and 6 of the main text. In the implementation of Eqs.~5 and 6, we adopt the approximation \cite{Souza2019} that Wannier orbitals are perfectly localized on atomic sites and spin matrix is half of Pauli matrix in the space spanned by Wannier orbitals. We use a constant broadening parameter $\eta=25~\rm{meV}$ for the results presented. The corresponding constant electron momentum relaxation time $\tau=\hbar/2\eta=13~\text{fs}$.  Since the critical N\'eel temperature of bilayer \ch{CrI3} is around 40 Kelvin, we adopt a low temperature $k_BT=3~\rm{meV}$.   \\

	\section{Symmetry-constrained forms of spin-orbit torque}
	\label{app:sym}
	\begin{figure}[htbp]
		\includegraphics[width=1\columnwidth]{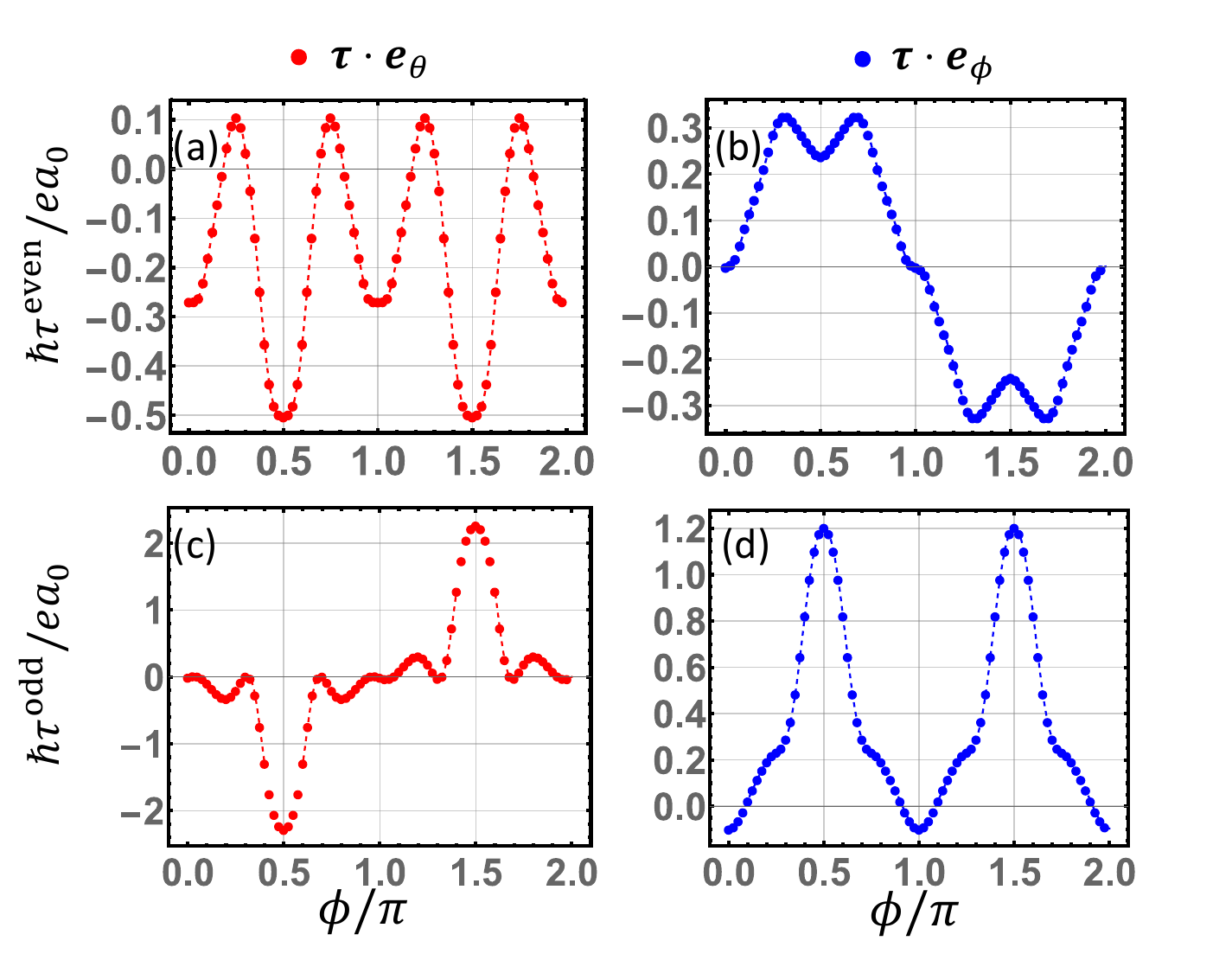}
		\caption{(Color online) Azimuthal angle ($\phi$) dependence of even and odd torkances at $\mu=0.05$~eV when $\theta=\pi/2$. Red  and blue circles denote torkance at the $\boldsymbol{e}_{\theta}$ and $\boldsymbol{e}_{\phi}$ direction respectively. Dashed lines show the fitted results based on the symmetry-constrained form Eq.~\ref{eq:fitting} up to $n=10$. The fitted even and odd torkances are $\boldsymbol{\tau}^{\rm even}=(0.32\sin\phi+0.03.\sin3\phi-0.05\sin5\phi) \boldsymbol{e}_{\phi}+(-0.17+0.1\cos2\phi-0.25\cos4\phi+0.02\cos6\phi+0.02\cos8\phi)\boldsymbol{e}_{\theta},\boldsymbol{\tau}^{\rm odd}=(0.38-0.53\cos2\phi+0.17\cos4\phi-0.13\cos6\phi)\boldsymbol{e}_{\phi}+(-1.0\sin\phi+0.6\sin3\phi-0.5\sin5\phi+0.3\sin7\phi)\boldsymbol{e}_{\theta}$. $a_0$ is the Bohr radius.
		}
		\label{Fig:higherorder}
	\end{figure}			
	In this section we provide the symmetry-constrained forms of the spin-orbit torque and fit the \textit{ab initio} results to these forms. In each layer of \ch{CrI3}, we only have one mirror plane $xz$. The time-reversal even and odd torkance under the applied field in $y$ direction are described by the symmetry-constrained expansion using a combination of trigonometric functions \cite{Garello2013}:
	\begin{widetext}
		\begin{eqnarray}
			\label{eq:fitting}
			\begin{split}
				\boldsymbol{\tau}^{\rm even}&=\sum_{m,n}[A_{mn}^{\rm even}\cos(2m\theta)\sin((2n+1)\phi)+B_{mn}^{\rm even}\sin(2m\theta)\sin(2n\phi)]\boldsymbol{e}_{\phi}\\
				&+[C_{mn}^{\rm even}\cos((2m+1)\theta)\cos((2n+1)\phi)+D_{mn}^{\rm even}\sin((2m+1)\theta)\cos(2n\phi)]\boldsymbol{e}_{\theta}
			\end{split},\\
			\begin{split}
				\boldsymbol{\tau}^{\rm odd}&=\sum_{m,n}[A_{mn}^{\rm odd}\cos((2m+1)\theta)\cos((2n+1)\phi)+B_{mn}^{\rm odd}\sin((2m+1)\theta)\cos(2n\phi)]\boldsymbol{e}_{\phi}\\
				&+[C_{mn}^{\rm odd}\cos(2m\theta)\sin((2n+1)\phi)+D_{mn}^{\rm odd}\sin(2m\theta)\sin(2n\phi)]\boldsymbol{e}_{\theta}
			\end{split},
		\end{eqnarray}
	\end{widetext}
	where $m(n)=0,1,2,...$. Note that coefficients $A, B, C, D$ are related since we need to ensure that the torque is independent of angle $\phi$ when $\theta=0, \pi$. We can immediately find that the conventional dampinglike and fieldlike forms of the torkance correspond to the lowest order contributions:
	\begin{widetext}
		\begin{eqnarray}
			&\boldsymbol{\tau}^{\rm even}=A_{00}^{\rm even}\sin\phi \boldsymbol{e}_{\phi}-A_{00}^{\rm even}\cos\theta\cos\phi \boldsymbol{e}_{\theta}+D_{00}^{\rm even}\sin\theta\boldsymbol{e}_{\theta}=\tau^{\rm even}\boldsymbol{m}\times(\boldsymbol{m}\times(p_x,0,p_z)),\\
			&\boldsymbol{\tau}^{\rm odd}=A_{00}^{\rm odd}\cos\theta\cos\phi\boldsymbol{e}_{\phi}+A_{00}^{\rm odd}\sin\phi\boldsymbol{e}_{\theta}+B_{00}^{\rm odd}\sin\theta\boldsymbol{e}_{\phi}=\tau^{\rm odd}\boldsymbol{m}\times(p_x,0,p_z).
		\end{eqnarray} 	
		
		The coefficients constraint $C_{00}^{\rm even,odd}=\mp A_{00}^{\rm even,odd}$ comes from the additional requirement that the torque must be independent of angle $\phi$ at the pole:
		
		\begin{eqnarray}
			&\boldsymbol{\tau}^{\rm even}(\theta\rightarrow0)=(-A_{00}^{\rm even}\sin^2\phi+C_{00}^{\rm even}\cos^2\phi,A_{00}^{\rm even}\sin\phi\cos\phi+C_{00}^{\rm even}\sin\phi\cos\phi,0),\\
			&\boldsymbol{\tau}^{\rm odd}(\theta\rightarrow0)=(-A_{00}^{\rm odd}\sin\phi\cos\phi+C_{00}^{\rm odd}\sin\phi\cos\phi,A_{00}^{\rm odd}\cos^2\phi+C_{00}^{\rm odd}\sin^2\phi,0).	
		\end{eqnarray}
	\end{widetext} 
	The unconventional symmetry direction $(p_x,0,p_z)$ is a consequence of the absence of mirror symmetry in both $xy$ and $yz$ planes. Fig.~\ref{Fig:higherorder} clearly shows the substantial higher-order contributions to both even and odd torkances. These higher order terms complicate the global torque sphere described in the main text.
	Note that we can also obtain the angular dependence of spin-orbit torque using a basis composed of orthogonal vector spherical harmonics \cite{Belashchenko2020} which takes care of the vector form automatically.
	
	\section{Spin-orbit torque in the pure N\'eel space}
	\label{app:Neel}
	\begin{figure}[htbp]
		\includegraphics[width=1\columnwidth]{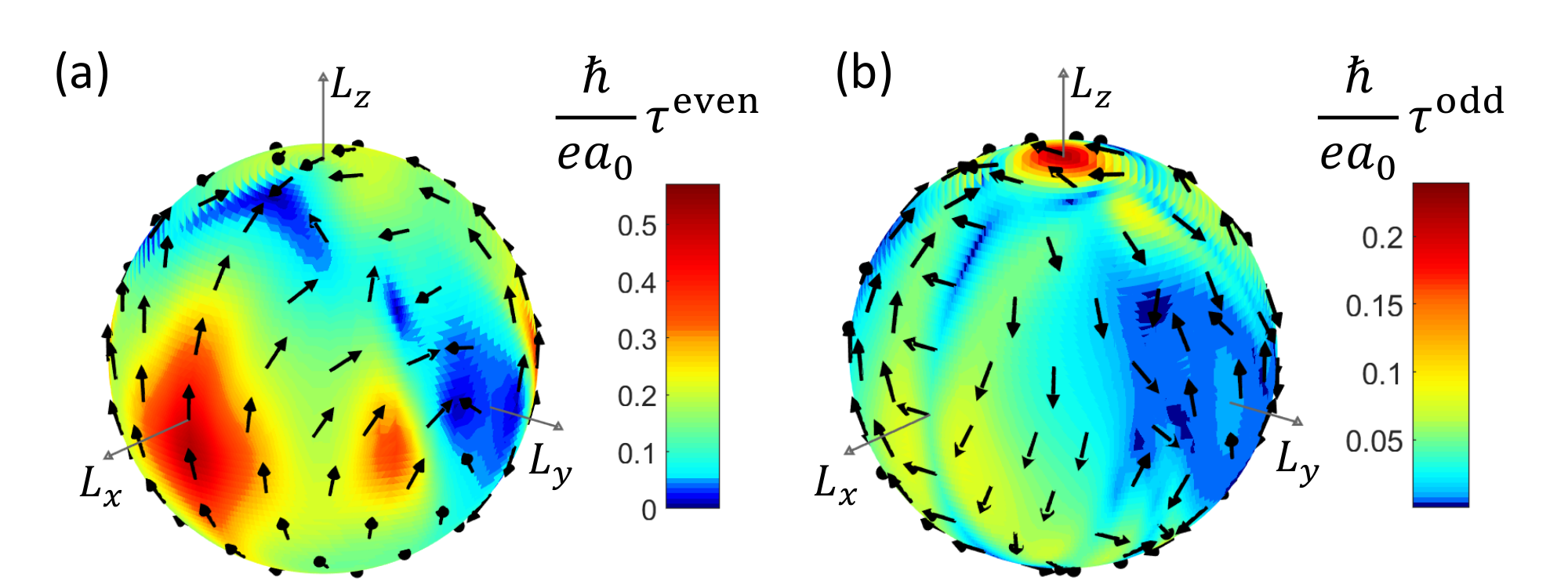}
		\caption{(Color online) Angular dependence of the dampinglike (a) and fieldlike (b) torkance on the N\'eel order direction $(\theta,\phi)$ for one layer of bilayer \ch{CrI3} under an external electric field along the $\hat{y}$ direction at Fermi level $\mu=0.06~\rm{eV}$. The arrow(color) on the sphere indicates the direction(magnitude) of the torkance under the given magnetization direction. We use $k_BT=3~\rm{meV}, \eta=30~\rm{meV}$ in the calculations. $a_0$ is the Bohr radius.
		}
		\label{fig:torque}
	\end{figure}	
	Here we present our first-principle results of spin-orbit torque in the pure N\'eel space, i.e., $\hat{\boldsymbol{m}}^{\rm A}=-\hat{\boldsymbol{m}}^{\rm B}$. In this case, the invariance under inversion+time reversal relates the torkance on the magnetic sublattices: The time-reversal even (dampinglike) torque is staggered while the time-reversal odd (fieldlike) torque is uniform. Figure ~\ref{fig:torque} summarizes our numerical results for $\mu=60$~meV above the conduction band edge. The results show similar features compared to the torkance in $N$-space shown in the main text, with fixed points in the $xz$ plane.  However, knowledge of the torkances in $L$-space is not sufficient for determining the spin dynamics since the anisotropy term immediately drives the system out of the pure N\'eel space. Note that the N\'eel space state is the same as the ${\bf N}$-space state at the $z$ and $x$ axes.

\end{document}